\documentclass[preprintnumbers,prd, twocolumn, 10pt, amsmath,amssymb,a4page, latexsym,array,enumerate,letter,numbers]{revtex4}
\usepackage{graphicx}
\usepackage{dcolumn}
\usepackage{bm}
\usepackage{amssymb}
\usepackage{epsfig}
\usepackage{slashed}
\usepackage{color}
\usepackage{calligra}
\usepackage[
       colorlinks=true,
      filecolor=black,
       anchorcolor=blue,
      linkcolor=blue,
      citecolor=cyan,
      urlcolor=cyan, 
       linktocpage=true,
        plainpages=false,
        breaklinks=true,
            pdfstartview=FitH
           ]{hyperref}
           
\def\cb{ \color[rgb]{0.00,0.00,0.00}} 

\newcommand{\nn}{\nonumber}

\def\MP{M_{\text{P}}}
\def\d{\rm d}
\def\a{\alpha}

\def\d{\rm d}
\def\a{\alpha}
\def\b{\beta}
\def\ns{N_p}
\def\as{{\xi}}
\def\T{{T}}

\begin{document}

\title{Notes on Quantum Corrections of Swampland and \\
Trans-Planckian Censorship Conjectures}

\author{Sichun Sun$^{1,2}$} \email[Corresponding author: ]{Sun.Sichun@roma1.infn.it}
\author{Yun-Long Zhang$^{3,4,5}$}\email[Corresponding author: ]{zhangyunlong@nao.cas.cn}
\affiliation{$^1$School of Physics, Beijing Institute of Technology, Haidian District, Beijing 100081, China,}
\affiliation{$^2$Department of Physics and INFN, Sapienza University of Rome, Rome I-00185, Italy}
\affiliation{$^3$National Astronomy Observatories, Chinese Academy of Science, Beijing, 100101, China}
\affiliation{$^4$School of Fundamental Physics and Mathematical Sciences, Hangzhou Institute for Advanced Study, University of Chinese Academy of Sciences, Hangzhou 310024, China}
\affiliation{$^5$Center for Gravitational Physics, Yukawa Institute for Theoretical Physics, Kyoto University, Sakyo-ku, Kyoto 606-8502, Japan}
  
\begin{abstract}We discuss the de Sitter swampland and trans-Planckian censorship conjectures (TCC) from the entropy bound with quantum corrections, namely quantum version of Bousso's bound and energy conditions. We include the typical contributions from the entanglement entropy in quasi-de Sitter spacetime. The TCC is not much corrected, whereas the bounds on de Sitter swampland conjecture from energy conditions can be relaxed due to quantum corrections.
\end{abstract}

\preprint{YITP-19-128}
\maketitle

\allowdisplaybreaks
 
 \section{Introduction}
 
The discussion around the allowed low-energy effective theories has been an interesting path to study the quantum gravity in ultraviolet(UV). Especially in the standard cosmology, early universe has a period of vacuum dominated time called inflation with quasi de Sitter(dS) geometry \cite{Guth:1980zm,Lyth:1998xn}. Since the inflation, as well as the dark energy dominated late universe geometry both imply that at least an approximate de Sitter vacuum exists in nature, it is important to construct such a vacuum/quasi vacuum in possible candidates of quantum gravity.  
The de Sitter swampland conjecture has been proposed \cite{Obied:2018sgi}, and originally stated that any vacuum has to satisfy $|\nabla V|\geq c\, V $ for constant $c \sim \mathcal{O} (1) $ to have a stringy construction, to forbid meta stable de Sitter vacuum. 
{\cb
In \cite{Ooguri:2018wrx}, the de Sitter swampland conjecture has been further refined as either $|\nabla V|/V \geq {\cal O}(1)\, $ or min$(\nabla_i \nabla_j V)<- {\cal O}(1)$. Although there are some counter-examples at scalar potential maxima that   $|\nabla V|/V\ll {\cal O}(1)$, the second condition on the minimum eigenvalue of the Hessian $\nabla_i \nabla_j V$ can still be satisfied. 
}
This conjecture, if being correct, basically states that string theory as the UV theory can not give rise to the standard $\Lambda$CDM universe.

Nevertheless, it is still worth exploring more that what kinds of low energy theories can arise from string theory and what might be the low energy theory for a consistent UV quantum gravity to resolve this sharpened contradiction.

 
More recently, a connection from the bounds on inflationary fluctuation to the de Sitter (dS) swampland conjecture has been proposed, namely trans-Planckian censorship conjecture (TCC) \cite{Bedroya:2019snp}. It comes from the statement that quantum fluctuations should remain quantum and never exit the Hubble horizon and freeze during inflation. This leads to the bound on the inflationary scale,
  \begin{align}
 H \lesssim \MP \,  e^{-N},
 \end{align}
where $\MP $ is the reduced Planck mass and $N$ is the $e$-folding number. 
This further leads to very small tensor-to-scalar ratio $r\lesssim 10^{-30}$ \cite{Bedroya:2019tba}, which is below current and near future observation reaches. If considering more general set-up, a  higher upper bound on the inflationary Hubble expansion rate $H_{\text{inf}} \lesssim \MP T_0/T_{\text{rh}}$ was proposed in \cite{Mizuno:2019bxy}, where $T_{\text{rh}}$ is the reheating temperature and $T_0$ is the photon temperature today.
In the lowest reheating temperature required for big bang nucleosynthesis, the bound on tensor-to-scalar ratio can be relaxed to be $r \lesssim 10^{-8}$. 
A series of relevant discussions can be found in \cite{Brahma:2019unn, Brahma:2019vpl, Kadota:2019dol,Schmitz:2019uti, Montero:2019ekk,  Cai:2019igo, Saito:2019tkc}. 
More interestingly, the dS swampland and TCC can also be derived from the entropy bound \cite{Ooguri:2018wrx,Kehagias:2019iem, Cai:2019dzj,Bonnefoy:2019nzv,Seo:2019wsh}.

 
In this note we discuss how the quantum corrections from entanglement contribution may relax both conjectures, and hope to shed light on future discussions incorporating the quantum effects in gravity.  It is known that \cite{Maldacena:2012xp} in four dimensions de Sitter space background with fluctuations the leading UV-divergent term of entanglement entropy gives rise to the well known area contributions from the surface particle entangled. The UV-finite piece of entropy contains the long range correlations that exit the Hubble horizon and freeze. For the logarithmic corrections, the UV-divergent part is proportional to $\log\left(\epsilon {H}\right)$ and the finite logarithmic piece can be sub-leading.   This logarithmic UV-divergent part is UV cut-off dependent and renormalizes the corresponding finite part. These added quantum effects can correct both Bousso's entropy bounds and null energy condition \cite{Bekenstein:1980jp, Bekenstein:1993dz, Brustein:1999md, Brustein:2001di}, then further modify dS swampland conjecture and TCC. 

The following part of this note is organized as below.
In section \ref{SecEE}, we derive the TCC from the entropy bound with quantum corrections. In section \ref{SecQNEC}, we consider the quantum null energy condition with corrections which leads to the modified dS swampland conjecture.
We conclude and discuss the result in section \ref{SecCon}.

\section{Quantum Corrections to the Entropy Bound} 
\label{SecEE}

On the entropy $S(R)$  in a region of radius $R$, the quantum version of the entropy bound has been proposed as 
\begin{align}\label{entropy1}
S(R) \leq S_{tot} =\MP^2  R^2 + S_{out}(R), 
\end{align}
where $\MP^2  R^2$ is the usual area law of the Bekenstein entropy for the surface, and $S_{out}(R)$ is the contribution from the entanglement entropy on one side of the surface, which contains UV divergences and can be regularized 
via the renormalization procedure. We refer to the related equations in \cite{Bousso:2018bli,Bousso:2015mna,Strominger:2003br} for more detailed discussions.

For the de Sitter case, according to the calculation in  \cite{Maldacena:2012xp}, one typical contribution to the entanglement entropy $S_{out}(R)$  is proportional to $ R^2 H^2+\cdots$, where $R$ is the superhorizon scale. 
 The entanglement entropy is contributed by the modes on the expanding de-Sitter background in four dimensions. Those modes are frozen after exiting the horizon.
There are also logarithmic contributions, which come from both local and long-range interactions. The local contribution is 
\begin{align}\label{EEdS}
S_{\log}(R)\simeq (\tilde \a  + \tilde \b  R^2 H^2) \log\left( \epsilon{H} \right)+\cdots,
\end{align}
where $\epsilon$ is the UV cutoff.
The long range contribution is the logarithmic term $\sim \log\left( R{H} \right)$ \cite{Maldacena:2012xp}.
During the inflationary era, if taking the UV cutoff $\epsilon\sim \MP^{-1}$, then in equation \eqref{EEdS} $\log\left( {\epsilon}{H}\right)\sim \log\left(\frac{H}{\MP}\right)$ can be approximated as a negative constant of order $\mathcal{O}(10)$. 
Some recent calculation for the corrections to de Sitter geometry entropy $\MP^2  R^2 $ from holography can also be found in \cite{Tetradis:2019vjn,Geng:2019bnn}, and they are also consistent with these typical terms. 

Though for more general fields in quasi de Sitter spacetime, these coefficients need to be calculated case by case, there is strong evidence that the divergence in $S_{out}(R)$ can be cancelled by the corrections to the geometry entropy $\MP^2  R^2 $ in \eqref{entropy1} \cite{Bousso:2015mna}.  
Thus, in the following estimation, we simply consider a few typical terms of the entanglement entropy 
as in \cite{Maldacena:2012xp}, 
\begin{align}\label{quantum}
S_{out}(R) \sim  (\a + \b  R^2 H^2) +\cdots,
\end{align}
which are UV cut-off independent, and contain the long range information.
Here $\alpha$ and $\beta$ are undetermined constants of order one, and $R$ can be a general scale. We will use the terms in \eqref{quantum} and consider their corrections to the entropy bound and energy conditions in the following derivations.

\subsection{From entropy bound to the TCC}
Considering $\ns$ pieces of relativistic particles at temperature ${\T}$, their energy and entropy are given by \cite{Kehagias:2019iem}
\begin{align}
{E}  ={\as} \ns R^3 {\T}^4, \qquad 
S(R) = \frac{4}{3} \frac{{E}}{{\T}}, \qquad {\as}\equiv \frac{4\pi^3}{45}.
\end{align}
The radius $R$ need to be larger than the Schwarzschild radius of a black hole with the same energy
\begin{align}
R \gtrsim R_s={E}/\MP^2 \quad \Rightarrow \quad
R \lesssim R_{m}\equiv \frac{{\MP}}{{\T}^2 \sqrt{{\as}\ns}},
\end{align}
with $R_{m}$ being the typical maximum radius, for the given temperature $T$ \cite{Dvali:2007wp}. 
{\cb
Therefore, the maximum entropy in the sphere is%
\begin{align}
S(R_m)=\frac{4 }{3}{\as} \ns R_m^3 {\T}^3 =\frac{4 }{3 \sqrt{{\as}\ns}}  \frac{\MP^3}{ {\T}^3 }\, .
\end{align}
}

From \eqref{entropy1} and \eqref{quantum}, we have
\begin{align}\label{inequality0}
S(R_m) \lesssim  \MP^2 R_m^2 +
(\a + 4\pi^2\b  R_m^2 {\T}^2), 
\end{align}
where $T_H\equiv \frac{H}{2\pi} \leq {\T}$ was used to replace $H$ in $\beta R_m^2 H^2\leq \beta R_m^2 (2\pi T)^2$,
which is guaranteed for $\beta>0$. For the case of $\beta<0$, the inequality is a bit arbitrary.
While considering the smallness of the quantum correction, we can simply take the oder estimation that $ {\T}\simeq \frac{H}{2\pi} $.
Considering $\alpha, \beta\ll n_s$ and ${\as}\sim \mathcal{O}(1)$,
and taking the equal sign in \eqref{inequality0}, to the sub-leading order we reach two solutions
\begin{align}\label{tems}
 \frac{{\T_1}}{\MP}& \simeq \frac{3}{4}\sqrt{\frac{{\as}}{\ns}} 
 \left(1+\frac{9\pi^2 {\as}}{4}\frac{ \b }{\ns}+\frac{81{\as}^3}{256}\frac{\a }{\ns}\right),\\
\frac{{\T_2}}{\MP}  & \simeq\frac{1}{3\pi^2} \sqrt{\frac{\ns}{{\as}}} \left(\frac{1}{\b}-\frac{\alpha}{\beta^4}\frac{\ns^2}{36\pi^6}
+\frac{\alpha}{\beta^3}\frac{{\as}\,{\ns}}{8\pi^4}\right)\, \label{tems1}.
\end{align}
We will take the first solution in \eqref{tems}, that higher orders have been omitted,
and the inequality \eqref{inequality0} implies $T\lesssim T_1 $.
The second solution in \eqref{tems1} has been expended for the case $\a\ll\b$, which is not physical any more.
It may only work for some special cases of $\alpha, \beta$ and provide an exception from the bound, although we will not pay attention to it in the following.

{\cb
From its relation to the effective cut-off and field $\Delta \phi$ transferred, the number of species satisfy, $\ns > e^{2 c \sqrt{\epsilon}N}$ with $c$ being a positive number of order one, and ${\epsilon}\equiv-\dot{H}/H^2$ \cite{Kehagias:2019iem,Cai:2019dzj,Seo:2019wsh}.
Again considering $T_H\equiv \frac{H}{2\pi} \leq {\T} \lesssim \T_1$,  we can recover the (refined) TCC bounds \cite{Cai:2019dzj,Seo:2019wsh} from the first solution in \eqref{tems} with positive $\beta$, such that
\begin{align}
 \frac{H}{\MP} \lesssim \frac{3\pi }{2}\frac{\sqrt{{\as}}}{e^{c \sqrt{\epsilon } N} } \left(1+\frac{9\pi^2 {\as}}{4}\frac{ \b }{\ns}+\frac{81{\as}^3}{256}\frac{\a }{\ns}\right).
\end{align}
}
We can see here for this condition that the quantum correction to the TCC bound is quite small as  $\alpha, \beta\ll n_s$. In the following section, it will be interesting to see that the bounds on dS swampland conjecture can be corrected.
with a non vanishing $\beta$.


\section{Quantum corrections to energy conditions}
\label{SecQNEC}

We consider the  $4$-dimensional effective theories which are compactified from the $D$-dimensional spacetime. 
It has been shown in \cite{Obied:2018sgi} that the strong energy condition leads to 
\begin{align}\label{SEC}
\frac{|\nabla{V}|}{V}\geq  \lambda_{\text{SEC} } , \qquad  \lambda_{\text{SEC}}=\sqrt{\frac{2(D-2)}{D-4}}\, ,
\end{align}
and null energy condition leads to 
\begin{align}\label{NEC}
\frac{|\nabla{V}|}{V}\geq  \lambda_{\text{NEC}}  \qquad  \lambda_{\text{NEC}}=\sqrt{\frac{2(D-4)}{D-2}} \, .
\end{align}
Both of these bounds are greater than the bound from TCC \cite{Bedroya:2019snp} 
for the asymptotic region of the moduli space 
\begin{align}\label{TCC}
\Big(\frac{|\nabla{V}|}{V}\Big)_{\infty}\geq  \lambda_{\text{TCC}} , \qquad  \lambda_{\text{TCC}}=\sqrt{\frac{2}{3}} \, .
\end{align}
In this section, we consider the  quantum corrections of null energy condition from \cite{Bousso:2018bli} which is derived from quantum focusing conjecture (QFC) \cite{Bousso:2015mna}:
\begin{align}
T_{MN} k^M k^N \geq \frac{\hbar}{2\pi} S_{out}''.
\end{align}
 {\cb
Here $k^M$ is the null vector orthogonal to the surface  with embedding function $X^\mu$ which divides the
spacetime into two parts. $S_{out}$ can be taken as the von Neumann entropy on one side of the surface and we will work in the unit of $\hbar=1$.  The double-prime is defined via the second functional derivative with respect to the deformation of the surface in the direction of $k^M$  \cite{Bousso:2018bli, Leichenauer:2018obf}, $\frac{\delta^2 S_{out}}{\delta X^k(y) \delta X^k(y')}=S''_{out}\delta^2(y-y')$. Alternatively in the integration formula, one has $ S''_{out}\equiv \text{limit}_{{\cal A} \to 0} \frac{1}{{\cal A} } \frac{d^{2} S_{out}(\lambda)}{d \lambda^{2}} $, where ${\cal A}$ is the small area element   \cite{Bousso:2015wca, Balakrishnan:2017bjg}.
For our case with $S_{out}$ in Eq. \eqref{quantum} as an order estimation and the deformation around the small area element ${\cal A} $, we can paramatrerize $S_{out}(\lambda)\sim \alpha+ \beta (1+\lambda)^2 {\cal A} H^2$ where  $\lambda$ is the affine parameter. Thus, we have $S''_{out}\sim 2\beta H^2$ and it will be used in the following derivations.
}



\subsection{Strong Energy Condition}

As a warm up of the derivation, we follow the derivation of the strong energy condition in \cite{Obied:2018sgi}.
 Consider the $D$-dimensional metric
\begin{align}
{\d}s^2 = \Omega(y,t)^2 (-{\d}t^2+ a(t)^2 {\d}x^2_{i}) +   {g}_{mn} {\d}y^m {\d}y^n,
\end{align}
where $x_i=x_1,x_2, x_3$ and $y^m$ being the $D-4$ dimensional indices of the internal space with metric $g_{mn}$ and warp factor $\Omega$. 

We choose the manifold such that $\kappa_4=\kappa_D$, which implies
\begin{align}\label{unit1}
\int d y^{D-4} \frac{\sqrt{g}}{\Omega^{2}}=1,\quad g\equiv \text{det}\, g_{mn}\,,
\end{align}
We can tune the initial conditions to make $\dot{\Omega}=\dot{g}_{mn}=0$.
Then taking the time derivatives of \eqref{unit1}, we have
\begin{align}
\int d y^{D-4} \frac{\sqrt{g}}{\Omega^{2}}\left(2\frac{\ddot{\Omega}}{\Omega}+\frac{1}{2}g^{mn}\ddot{g}_{mn}\right)=1.
\end{align}


The strong energy condition in general dimension can be written as $(T_{MN}-\frac{T^D_{~D}}{D-2} g_{MN}) v^M v^N=R_{MN}^{(D)} v^M v^N\geq 0$, for any timelike vector $v^M$ \cite{Maeda:2018hqu}.
As a practice below, we also add the quantum correction to the strong energy condition. Here we take $v^N=({1, 0, ...})$, and consider the quantum corrections as below

\begin{align}\label{ec1}
&  \left(T_{tt}  - \frac{g_{tt}}{D-2} T^D_{~D} \right) =R_{tt}^{(D)}
= - 3\frac{\ddot {a}}{a} -3\frac{\ddot \Omega}{\Omega} \nonumber\\
&+\left[-\frac{1}{2}  {g}^{mn} {\ddot{  {g}}_{mn}} +\frac{1}{3\Omega^2}  \nabla^m \nabla_m (\Omega^3) \right]\geq 
\frac{ S''_{out}}{2\pi} \, .
\end{align}
Notice here we only consider the entanglement entropy corrections on the four large dimensions, with possible modes localized on the 4-dimension surface. As an order estimation, we simply use entanglement entropy on a de sitter background result in Eq.(\ref{quantum}),
and take $S''_{out}(R)\sim 2\beta H^2$.

For the 4-dimensional Friedmann equation without kinetic energy, $H^2 \equiv \frac{\dot{a}^2}{a^2}=\frac{\ddot a}{a }$.
Multiplying \eqref{ec1} by $\Omega^{2}$ and integrating over the compact manifold, we end up with
\begin{align}
\left(3+ \frac{\b}{\pi}\right) H^2 \leq  -\int {\d}y^{D-4} \Omega^2 \sqrt{ {g}} \frac{\ddot \Omega}{\Omega}\, .
\end{align}
In order to relate this to a bound on the potential, we need to analyze the 4-d effective action
\begin{align}
S=\int dx^4\sqrt{g}\left[ \frac{\MP^2}{2} \left(R^{(4)} +  \dot{\phi}^2\right) -V(\phi,\Phi)+...\right].
\end{align}
At the point where $\dot{\phi}=0$ and $\dot{\Phi}=0$, the equations of motion in 4-dimension are
\begin{align}
-\MP^2 \ddot{\phi}&=\partial_\phi V(\phi,\Phi)\nn\\
3\MP^2H^2&=V(\phi,\Phi)\,.
\end{align}

Solving the Einstein equations as in  \cite{Obied:2018sgi}, 
we now can derive something similar to the dS swampland conjecture:
\begin{align}\label{lambdaS}
 \frac{ |\nabla V|}{V}  \geq  \sqrt{\frac{2(D-2)}{D-4}}\left(1+\frac{{\b}}{3\pi}\right)
\equiv {\lambda_{QSEC}}.
\end{align}
When taking $\beta \to 0$, we recover the bound $\lambda_{SEC}$ from strong energy condition in \eqref{SEC}.
The bound will become stringent for the positive $\beta$, or loose for a negative $\beta$. Significantly, if $\beta\leq-3\pi$, there is no bound on this anymore.
Thus, the quantum correction turns out to be important to the bounds on dS swampland conjecture.

\subsection{Quantum Null Energy Condition}

For the null energy condition, 
consider the $D$-dimensional metric 
\begin{align}
{\d}s^2 &= \Omega(y,t)^2 (-{\d}t^2+ a(t) {\d}x^2_{i}) + \Omega(y,t)^{- 2 \gamma}   {h}_{mn} {\d}y^m {\d}y^n,\nonumber\\
 \gamma &\equiv \frac{d}{D-d-2}=\frac{4}{D-6}\, .
\end{align}
 That we have put in $d=4$.
Following \cite{Obied:2018sgi}, we tune the initial conditions $\dot{\Omega}=\dot{h}_{mn}=0$ and fix the frame via
\begin{align}\label{unit2}
\int d y^{D-4} {\sqrt{h}}\,{\Omega^{-2(\gamma+1)}}=1,\quad h\equiv \text{det}\, h_{mn}\,.
\end{align}

Considering the null vector $k^t k^t=\Omega^{-2}$ and $k^m k^n=\frac{\Omega^{2\gamma}}{D-4} h^{mn}$,  we have similar derivation and arrive at
\begin{align}
&T_{MN} \langle k^M k^N\rangle =  R_{MN} \langle k^M k^N\rangle \nonumber\\
&=\,\frac{1}{\Omega^2}R_{tt}^{(D)}+ \frac{\Omega^{2\gamma}}{D-4} R_{mn}^{(D)} h^{mn} \geq  \frac{S''_{out}}{2\pi} \, .
\end{align}
Again, we use Eq.(\ref{quantum}) and take $S''_{out}(R)\sim 2\beta H^2$.

Multiplying by $\Omega^{-2\gamma}$ and integrating over the compact manifold sphere, 
we end up with
\begin{align}
 \left(3+ \frac{ C{\b}}{\pi} \right) H^2\leq
-\int {\d}y^{D-4} \frac{\sqrt{  h}}{\Omega^{2(\gamma+1)}}  \frac{D-2}{D-4} \frac{\ddot \Omega}{\Omega}\, .
\end{align}
The condition \eqref{unit2} has been used, and the constant $C=\int d y^{D-4} {\sqrt{ h}}{\Omega^{-2\gamma}}>0$ comes from integrating over the compact manifold.
Solving the Einstein equation for the $\Omega$, we now can derive something similar to the {\cb dS swampland} conjecture,
\begin{align}\label{lambdaN}
 \frac{ |\nabla V|}{V} \geq
 \sqrt{\frac{2(D-4)}{D-2}}\left(1+ \frac{ C{\b}}{3\pi}\right)
\equiv  {\lambda_{QNEC}}.
\end{align}
When $\beta \to 0$, we recover the bound from null energy condition in \eqref{NEC}. 
Again, if $\beta\leq-3\pi/C$, there is no bound on ${|\nabla V|}/{V}$ anymore, such that the de Sitter vacuum might be allowed.

\vspace{-0pt}
\section{Discussions}
\label{SecCon}

For a few more discussions on the swampland conditions relevant to the quantum effects, entropy bound and energy conditions, see e.g. \cite{Vafa:2005ui,Ooguri:2006in, Brustein:2007hd, Agrawal:2018own, Dvali:2018fqu, Dasgupta:2018rtp,Han:2018yrk, Ooguri:2018wrx, Dvali:2018jhn, Cai:2018ebs, Palti:2019pca, Mizuno:2019pcm,Lust:2019zwm,Geng:2019zsx}.  
In view of \eqref{lambdaS} and \eqref{lambdaN}, for the case of $D=10$, it is interesting to see that both the bounds $\lambda_{QSEC}=2\sqrt{\frac{2}{3}}\left(1+\frac{\beta}{3\pi}\right)$ and $\lambda_{QNEC}=\sqrt{\frac{3}{2}}\left(1+\frac{C \beta }{3\pi}\right)$ are corrected due to the quantum corrections in \eqref{quantum}.
For a negative $\beta$, they might be comparable with $\lambda_{TCC}=\sqrt{2/3}$ from \eqref{TCC} in the asymptotic region.
Significantly, if we consider the holographic entanglement entropy for de Sitter spacetime in \cite{Tetradis:2019vjn}, which leads to $\b\equiv \tilde \b \log\left(\frac{H}{\MP}\right)\sim - \mathcal{O}(10)$, such that $\lambda_{QSEC}$ and $\lambda_{QNEC}$ can be negative, which lead to no bound on ${|\nabla V|}/{V}$ anymore. In view of this, it is not impossible to have the dS vacua in string theory from compactification with quantum effects, such that the tension between swampland conjecture and standard cosmology can be relaxed.

Thus, it is one step from classical condition to quantum interpretation, although the exact meaning supporting these bounds still needs further developments and we just naively incorporate the leading contributions here. The counting of entropy within the quantum focusing conjecture is also related to the recent discussions of the black hole entropy, as we are taking into account the entropy outside the horizon \cite{Almheiri:2019yqk}.  It would also be interesting to discuss these effects in the framework of the compactification constructions, if possible, 
see e.g. \cite{Dasgupta:2019vjn, Banks:2019oiz, Blumenhagen:2019vgj}. 
 
{\cb 
 Besides, the de Sitter swampland conjecture has been further refined as either $|\nabla V|/V \geq {\cal O}(1)\, $ or min$(\nabla_i \nabla_j V)<- {\cal O}(1)$ in \cite{Ooguri:2018wrx}, to include some counter examples. Although the second condition can not be derived from the null energy condition as in our section \ref{SecQNEC}, it can be related to the distance conjecture by using entropy bound and  it was refined further in \cite{Seo:2019wsh}. Thus, it is also meaningful to see how the second condition is affected by the quantum corrections in further studies.
} 


In summary, we study the dS swampland and trans-Planckian censorship conjectures from the entropy bound with quantum corrections. Especially, we consider the typical contributions from the entanglement entropy in de Sitter spacetime and discuss the quantum corrections to both conjectures.   The bound on TCC is found to be not much corrected, whereas the bounds on swampland conjecture can be relaxed. It is interesting to consider the entanglement entropy contributions, in the process of understanding the swampland and TCC in more details.

\vspace{-0pt}
\section*{Acknowledgement}
We thank S. Mizuno, S. Mukohyama, S. Pi, and H. J. Wang for helpful conversations.
S.\, Sun was supported by MIUR in Italy under Contract (No.PRIN 2015P5SBHT) and European Research Council (ERC) Ideas Advanced Grant (No.267985) \textquotedblleft DaMeSyFla";\, 
Y.\, -L.\, Zhang was supported by Grant-in-Aid for JSPS international research fellow (No.18F18315), and by National Natural Science Foundation of China (NSFC) grant (No.12005255).


\begin{thebibliography}{99}
 \scriptsize
\footnotesize
\bibitem{Guth:1980zm} 
  A.~H.~Guth,
{}``The Inflationary Universe: A Possible Solution to the Horizon and Flatness Problems,''
  \href{https://doi.org/10.1103/PhysRevD.23.347}{Phys.\ Rev.\ D {\bf 23}, 347 (1981)}
  [Adv.\ Ser.\ Astrophys.\ Cosmol.\  {\bf 3}, 139 (1987)].
\bibitem{Lyth:1998xn} 
  D.~H.~Lyth and A.~Riotto,
{}``Particle physics models of inflation and the cosmological density perturbation,''
  Phys.\ Rept.\  {\bf 314}, 1 (1999)
  [ \href{https://arxiv.org/abs/hep-ph/9807278}{hep-ph/9807278}].



\bibitem{Obied:2018sgi} 
  G.~Obied, H.~Ooguri, L.~Spodyneiko and C.~Vafa,
``de Sitter Space and the Swampland,''
  \href{https://arxiv.org/abs/1806.08362}{arXiv:1806.08362 [hep-th]}.


\bibitem{Bedroya:2019snp} 
  A.~Bedroya and C.~Vafa,
{}``Trans-Planckian Censorship and the Swampland,''
   \href{https://arxiv.org/abs/1909.11063}{arXiv:1909.11063 [hep-th]}.

\bibitem{Bedroya:2019tba} 
  A.~Bedroya, R.~Brandenberger, M.~Loverde and C.~Vafa,
``Trans-Planckian Censorship and Inflationary Cosmology,''
  \href{https://arxiv.org/abs/1909.11106}{arXiv:1909.11106 [hep-th]}.

 
\bibitem{Mizuno:2019bxy}
S.~Mizuno, S.~Mukohyama, S.~Pi and Y.~L.~Zhang,
Phys. Rev. D \textbf{102}, no.2, 021301 (2020)
[ \href{https://arxiv.org/abs/1910.02979}{arXiv:1910.02979 [astro-ph.CO]}].


  
  

  
  
  
 
\bibitem{Brahma:2019unn} 
  S.~Brahma,
  ``Trans-Planckian censorship, inflation and excited initial states for perturbations,''
  Phys.\ Rev.\ D {\bf 101}, no. 2, 023526 (2020)
  [\href{http://arxiv.org/abs/arXiv:1910.04741}{arXiv:1910.04741 [hep-th]}].

\bibitem{Schmitz:2019uti} 
  K.~Schmitz,
  ``Trans-Planckian Censorship and Inflation in Grand Unified Theories,''
  Phys.\ Lett.\ B {\bf 803}, 135317 (2020)
  [\href{http://arxiv.org/abs/arXiv:1910.08837}{arXiv:1910.08837 [hep-ph]}].
  
\bibitem{Kadota:2019dol} 
  K.~Kadota, C.~S.~Shin, T.~Terada and G.~Tumurtushaa,
``Trans-Planckian censorship and single-field inflation potential,''
  \href{http://arxiv.org/abs/arXiv:1910.09460}{arXiv:1910.09460 [hep-th]}.

\bibitem{Brahma:2019vpl} 
  S.~Brahma,
  ``Trans-Planckian censorship conjecture from the swampland distance conjecture,''
  Phys.\ Rev.\ D {\bf 101}, no. 4, 046013 (2020)
  [\href{http://arxiv.org/abs/arXiv:1910.12352}{arXiv:1910.12352 [hep-th]}].
  
  
  

  
  
  
  
  

  

  
\bibitem{Montero:2019ekk} 
  M.~Montero, T.~Van Riet and G.~Venken,
   ``Festina Lente: EFT Constraints from Charged Black Hole Evaporation in de Sitter,''
  JHEP {\bf 2001}, 039 (2020)
  [\href{https://arxiv.org/abs/1910.01648}{arXiv:1910.01648 [hep-th]}].
  
  
  
  
  
  
  
  
  
  
 


\bibitem{Cai:2019igo} 
  R.~G.~Cai and S.~J.~Wang,
 ``Mass bound for primordial black hole from trans-Planckian censorship conjecture,''
  Phys.\ Rev.\ D {\bf 101}, no. 4, 043508 (2020)
  [\href{http://arxiv.org/abs/arXiv:1910.07981}{arXiv:1910.07981 [astro-ph.CO]}].
  

\bibitem{Saito:2019tkc} 
  R.~Saito, S.~Shirai and M.~Yamazaki,
 ``Is the trans-Planckian censorship a swampland conjecture?,''
  Phys.\ Rev.\ D {\bf 101}, no. 4, 046022 (2020)
  [\href{http://arxiv.org/abs/arXiv:1911.10445}{arXiv:1911.10445 [hep-th]}].










\bibitem{Seo:2019wsh}
M.~S.~Seo,
``The entropic quasi-de Sitter instability time from the distance conjecture,''
Phys. Lett. B \textbf{807}, 135580 (2020)
[\href{https://arxiv.org/abs/1911.06441}{arXiv:1911.06441 [hep-th]}].
 

\bibitem{Kehagias:2019iem} 
  A.~Kehagias and A.~Riotto,
 ``A Note on the Swampland Distance Conjecture,''
  Fortsch.\ Phys.\  {\bf 68}, no. 1, 1900099 (2020)
  [ \href{https://arxiv.org/abs/1911.09050}{arXiv:1911.09050 [hep-th]}].

\bibitem{Cai:2019dzj} 
  R.~G.~Cai and S.~J.~Wang,
``A refined trans-Planckian censorship conjecture,''
  \href{https://arxiv.org/abs/1912.00607}{arXiv:1912.00607 [hep-th]}.
  
  


\bibitem{Bonnefoy:2019nzv} 
  Q.~Bonnefoy, L.~Ciambelli, D.~Lüst and S.~Lüst,
  ``Infinite Black Hole Entropies at Infinite Distances and Tower of States,''
  Nucl.\ Phys.\ B {\bf 958}, 115112 (2020)
  [\href{http://arxiv.org/abs/1912.07453}{arXiv:1912.07453 [hep-th]}].

\bibitem{Maldacena:2012xp} 
  J.~Maldacena and G.~L.~Pimentel,
``Entanglement entropy in de Sitter space,''
  JHEP {\bf 1302}, 038 (2013)
  [\href{https://arxiv.org/abs/1210.7244}{arXiv:1210.7244 [hep-th]}].
  

  
\bibitem{Bekenstein:1980jp} 
  J.~D.~Bekenstein,
``A Universal Upper Bound on the Entropy to Energy Ratio for Bounded Systems,''
  \href{https://doi.org/10.1103/PhysRevD.23.287}{Phys.\ Rev.\ D {\bf 23}, 287 (1981)}.
  
\bibitem{Bekenstein:1993dz}
  J.~D.~Bekenstein,
``Entropy bounds and black hole remnants,''
 Phys.\ Rev.\ D {\bf 49}, 1912 (1994)
  [\href{https://arxiv.org/abs/hgr-qc/9307035}{gr-qc/9307035}].
\bibitem{Brustein:1999md} 
  R.~Brustein and G.~Veneziano,
``A Causal entropy bound,''
  Phys.\ Rev.\ Lett.\  {\bf 84}, 5695 (2000)
  [\href{https://arxiv.org/abs/hep-th/9912055}{hep-th/9912055}].
\bibitem{Brustein:2001di} 
  R.~Brustein, S.~Foffa and G.~Veneziano,
``CFT, holography, and causal entropy bound,''
  Phys.\ Lett.\ B {\bf 507}, 270 (2001)
  [\href{https://arxiv.org/abs/hep-th/0101083}{hep-th/0101083}].


  \bibitem{Bousso:2018bli} 
  R.~Bousso,
``Black hole entropy and the Bekenstein bound,''
  \href{https://arxiv.org/abs/1810.01880}{arXiv:1810.01880 [hep-th]}.



\bibitem{Bousso:2015mna} 
  R.~Bousso, Z.~Fisher, S.~Leichenauer and A.~C.~Wall,
``Quantum focusing conjecture,''
  Phys.\ Rev.\ D {\bf 93}, no. 6, 064044 (2016)
  [\href{https://arxiv.org/abs/1506.02669}{arXiv:1506.02669 [hep-th]}].
  


\bibitem{Strominger:2003br} 
  A.~Strominger and D.~M.~Thompson,
``A Quantum Bousso bound,''
  Phys.\ Rev.\ D {\bf 70}, 044007 (2004)
  [\href{http://arxiv.org/abs/hep-th/0303067}{hep-th/0303067}].



\bibitem{Tetradis:2019vjn}
N.~Tetradis,
``Corrections to de Sitter entropy through holography,''
Phys. Lett. B \textbf{807}, 135552 (2020)
[\href{https://arxiv.org/abs/1910.10587}{arXiv:1910.10587 [hep-th]}].
  

\bibitem{Geng:2019bnn} 
  H.~Geng, S.~Grieninger and A.~Karch,
``Entropy, Entanglement and Swampland Bounds in DS/dS,''
  JHEP {\bf 1906}, 105 (2019)
  [\href{https://arxiv.org/abs/1904.02170}{arXiv:1904.02170 [hep-th]}].

 

\bibitem{Dvali:2007wp} 
  G.~Dvali and M.~Redi,
 ``Black Hole Bound on the Number of Species and Quantum Gravity at LHC,''
  Phys.\ Rev.\ D {\bf 77}, 045027 (2008)
  [\href{https://arxiv.org/abs/0710.4344}{arXiv:0710.4344 [hep-th]}].
 
\bibitem{Bousso:2015wca} 
  R.~Bousso, Z.~Fisher, J.~Koeller, S.~Leichenauer and A.~C.~Wall,
 Phys.\ Rev.\ D {\bf 93}, no. 2, 024017 (2016)
  [\href{https://arxiv.org/abs/1509.02542}{arXiv:1509.02542 [hep-th]}].
\bibitem{Balakrishnan:2017bjg} 
  S.~Balakrishnan, T.~Faulkner, Z.~U.~Khandker and H.~Wang,
  JHEP {\bf 1909}, 020 (2019)
  [\href{https://arxiv.org/abs/1706.09432}{arXiv:1706.09432 [hep-th]}].

\bibitem{Leichenauer:2018obf} 
  S.~Leichenauer, A.~Levine and A.~Shahbazi-Moghaddam,
   ``Energy density from second shape variations of the von Neumann entropy,''
  Phys.\ Rev.\ D {\bf 98}, no. 8, 086013 (2018)
  [\href{http://arxiv.org/abs/arXiv:1802.02584}{arXiv:1802.02584 [hep-th]}].
   

\bibitem{Maeda:2018hqu}
H.~Maeda and C.~Martinez,
 ``Energy conditions in arbitrary dimensions,''
PTEP \textbf{2020}, no.4, 043E02 (2020)
doi:10.1093/ptep/ptaa009
[\href{https://arxiv.org/abs/1810.02487}{arXiv:1810.02487 [gr-qc]}].
  

  
  
  
 
\bibitem{Vafa:2005ui} 
  C.~Vafa,
 ``The String landscape and the swampland,''
\href{https://arxiv.org/abs/hep-th/0509212}{hep-th/0509212}.
\bibitem{Ooguri:2006in} 
  H.~Ooguri and C.~Vafa,
 ``On the Geometry of the String Landscape and the Swampland,''
  Nucl.\ Phys.\ B {\bf 766}, 21 (2007)
  [\href{https://arxiv.org/abs/hep-th/0605264}{hep-th/0605264}].
\bibitem{Brustein:2007hd} 
  R.~Brustein,
 ``Cosmological entropy bounds,''
  Lect.\ Notes Phys.\  {\bf 737}, 619 (2008)
  [\href{https://arxiv.org/abs/hep-th/0702108}{hep-th/0702108}].
  
  
\bibitem{Agrawal:2018own} 
  P.~Agrawal, G.~Obied, P.~J.~Steinhardt and C.~Vafa,
``On the Cosmological Implications of the String Swampland,''
  Phys.\ Lett.\ B {\bf 784}, 271 (2018)
  [\href{https://arxiv.org/abs/1806.09718}{arXiv:1806.09718 [hep-th]}].
  
  
\bibitem{Dvali:2018fqu} 
  G.~Dvali and C.~Gomez,
``On Exclusion of Positive Cosmological Constant,''
  Fortsch.\ Phys.\  {\bf 67}, no. 1-2, 1800092 (2019)
  [\href{https://arxiv.org/abs/1806.10877}{arXiv:1806.10877 [hep-th]}].

  
  
  
\bibitem{Dasgupta:2018rtp} 
  K.~Dasgupta, M.~Emelin, E.~McDonough and R.~Tatar,
``Quantum Corrections and the de Sitter Swampland Conjecture,''
  JHEP {\bf 1901}, 145 (2019)
  [\href{http://arxiv.org/abs/arXiv:1808.07498}{arXiv:1808.07498 [hep-th]}].

\bibitem{Han:2018yrk} 
  C.~Han, S.~Pi and M.~Sasaki,
``Quintessence Saves Higgs Instability,''
  Phys.\ Lett.\ B {\bf 791}, 314 (2019)
  [\href{http://arxiv.org/abs/arXiv:1809.05507}{arXiv:1809.05507 [hep-ph]}].
  
\bibitem{Ooguri:2018wrx} 
  H.~Ooguri, E.~Palti, G.~Shiu and C.~Vafa,
``Distance and de Sitter Conjectures on the Swampland,''
  Phys.\ Lett.\ B {\bf 788}, 180 (2019)
  [\href{http://arxiv.org/abs/1810.05506}{arXiv:1810.05506 [hep-th]}].



\bibitem{Dvali:2018jhn} 
  G.~Dvali, C.~Gomez and S.~Zell,
``Quantum Breaking Bound on de Sitter and Swampland,''
  Fortsch.\ Phys.\  {\bf 67}, no. 1-2, 1800094 (2019)
  [\href{http://arxiv.org/abs/arXiv:1810.11002}{arXiv:1810.11002 [hep-th]}].
  
  
  
  
  
\bibitem{Cai:2018ebs} 
  R.~G.~Cai, S.~Khimphun, B.~H.~Lee, S.~Sun, G.~Tumurtushaa and Y.~L.~Zhang,
``Emergent Dark Universe and the Swampland Criteria,''
  Phys.\ Dark Univ.\  {\bf 26}, 100387 (2019)
  [\href{http://arxiv.org/abs/1812.11105}{arXiv:1812.11105 [hep-th]}].
  
  
  
  
\bibitem{Palti:2019pca} 
  E.~Palti,
``The Swampland: Introduction and Review,''
  Fortsch.\ Phys.\  {\bf 67}, no. 6, 1900037 (2019)
  [\href{http://arxiv.org/abs/1903.06239}{arXiv:1903.06239 [hep-th]}].
  
  
  
  
\bibitem{Mizuno:2019pcm} 
  S.~Mizuno, S.~Mukohyama, S.~Pi and Y.~L.~Zhang,
``Hyperbolic field space and swampland conjecture for DBI scalar,''
  JCAP {\bf 1909}, no. 09, 072 (2019)
  [\href{http://arxiv.org/abs/1905.10950}{arXiv:1905.10950 [hep-th]}].
  
  \bibitem{Lust:2019zwm} 
  D.~Lüst, E.~Palti and C.~Vafa,
``AdS and the Swampland,''
  Phys.\ Lett.\ B {\bf 797}, 134867 (2019)
  [\href{http://arxiv.org/abs/1906.05225}{arXiv:1906.05225 [hep-th]}].
  
  

\bibitem{Geng:2019zsx} 
  H.~Geng,
``Distance Conjecture and De-Sitter Quantum Gravity,''
  Phys.\ Lett.\ B {\bf 803}, 135327 (2020)
  [\href{http://arxiv.org/abs/1910.03594}{arXiv:1910.03594 [hep-th]}].


\bibitem{Almheiri:2019yqk}   
 A.~Almheiri, R.~Mahajan and J.~Maldacena,
 ``Islands outside the horizon,'' 
 \href{http://arxiv.org/abs/arXiv:1910.11077}{arXiv:1910.11077 [hep-th]}.
 
 

\bibitem{Blumenhagen:2019vgj} 
  R.~Blumenhagen, M.~Brinkmann and A.~Makridou,
 ``Quantum Log-Corrections to Swampland Conjectures,''
  JHEP {\bf 2002}, 064 (2020)
  [\href{http://arxiv.org/abs/1910.10185}{arXiv:1910.10185 [hep-th]}].
 
 
\bibitem{Banks:2019oiz} 
  T.~Banks,
``On the Limits of Effective Quantum Field Theory: Eternal Inflation, Landscapes, and Other Mythical Beasts,''
  \href{http://arxiv.org/abs/arXiv:1910.12817}{arXiv:1910.12817 [hep-th]}.

\bibitem{Dasgupta:2019vjn} 
  K.~Dasgupta, M.~Emelin, M.~M.~Faruk and R.~Tatar,
``How a four-dimensional de Sitter solution remains outside the swampland,''
  \href{http://arxiv.org/abs/arXiv:1911.02604}{arXiv:1911.02604 [hep-th]}.
  


 


  
\end{thebibliography}
\end{document}